\documentclass[]{aa}
\usepackage{graphics,psfig}

\newcommand{\bra}{Br$\alpha$}
\newcommand{\pfg}{Pf$\gamma$}
\newcommand{\hu}{Hu(14-6)}
\newcommand{\hi}{\ion{H}{i}}
\newcommand{\hei}{\ion{He}{i}}
\newcommand{\civ}{\ion{C}{iv}}
\newcommand{\siv}{\ion{Si}{iv}}
\newcommand{\nv}{\ion{N}{v}}
\begin{document}

\title{Hydrogen infrared recombination lines as a diagnostic tool for the geometry of the circumstellar material of hot stars\thanks{Based on observations with ISO, an ESA
project with instruments funded by ESA Member States (especially
the PI countries: France, Germany, the Netherlands and the United
Kingdom) and with the participation of ISAS and NASA}}
\author{A. Lenorzer\inst{1}
A. de Koter\inst{1},
L.B.F.M. Waters\inst{1,2}}
\institute{
Astronomical Institute 'Anton Pannekoek', University of Amsterdam,
Kruislaan 403, 1098 SJ Amsterdam, The Netherlands
\and
Instituut voor Sterrenkunde, K.U. Leuven, Celestijnenlaan 200B,
3001 Heverlee, Belgium}

\offprints{A. Lenorzer:\\ lenorzer@astro.uva.nl}

\date{received\dots, accepted\dots}

\authorrunning{A. Lenorzer et al.}
\titlerunning{Hydrogen infrared recombination lines as diagnostic 
tool for hot star wind geometry}

\abstract{We have analysed the infrared hydrogen recombination
lines of a sample of well studied hot massive stars observed with
the Infrared Space Observatory. Our sample contains stars from
several classes of objects, whose circumstellar environment is
believed to be dominated by an ionized stellar wind (the Luminous
Blue Variables) or by a dense disk-like geometry (Be
stars and B[e] stars). We show that hydrogen infrared
recombination lines can be used as a diagnostic tool to constrain
the geometry of the ionized circumstellar  material. The line
strengths are sensitive to the density of the emitting gas. 
High densities result in optically thick lines for which line strengths are 
only dependent on the emitting surface. 
Low density gas produces optically thin lines which may be characterized by
Menzel case B  
recombination.
The ISO observations show
that stellar winds are dominated by optically thin \hi\
recombination lines, while disks are dominated by optically thick
lines. Disks and winds are well separated in a diagnostic diagram
using the \hu/\bra\ and the \hu/\pfg\ line
flux ratios. This diagnostic tool is useful to constrain the
nature of hot star environments in case they are highly obscured,
for instance while they are still embedded in their
natal molecular cloud.}

\maketitle


\section{Introduction}

Hot massive stars are characterized by the presence of ionized
circumstellar matter, which is in the majority of cases due to a
stellar wind. The presence of this ionized gas is easily detected
at ultraviolet wavelengths through P~Cygni profiles of resonance
lines of e.g. \civ, \siv\ and \nv, while at optical and infrared
wavelengths free-bound and free-free continuum emission as well as
\hi\ recombination line radiation can be observed. Both the strength
and the shape of the lines formed in the circumstellar gas are
sensitive to the densities, velocities and the geometry of the
gas. For instance, the P~Cygni profiles in H$\alpha$ seen in hot
stars with dense winds are indicative of a roughly spherical
outflow whose terminal velocity can be derived from the blue edge
of the absorption part of the profile (e.g. Castor \& Lamers
1979). The ionized disks surrounding Be stars show a
characteristic double-peaked emission line structure, and can be
explained by almost Keplerian rotation with hardly any radial
expansion (Struve 1933; Limber \& Marlborough 1968). The situation is less
clear in the case of the so-called B[e] stars, partly due to the
mixed nature of the stars in this group (see Lamers et al. 1998).
However, most of these stars are believed to be surrounded by a
flattened envelope.

While the wind diagnostics at UV and optical wavelengths are well
studied and an extensive literature exists, the situation is less
well documented at infrared wavelengths. This is mostly due to the
unavailability, until recently, of high quality infrared spectra.
As infrared instrumentation becomes more sensitive, the stellar
population of highly obscured regions becomes
accessible, such as the galactic centre or star
forming regions. Since traditional wind diagnostics are no longer
usable in these regions, infrared observations must be used to
characterize the stellar populations and their circumstellar
matter. 

In this \emph{Letter}, we present a simple diagnostic
diagram which allows one to constrain the geometry of the ionized
part of the circumstellar envelope of hot stars by means of
measuring line flux ratios in a few well-chosen \hi\ recombination
lines. We calibrate this new diagnostic tool using
well-studied, and \emph{optically bright} hot stars observed with
the Short Wavelength Spectrometer (SWS, de Graauw et al. 1996) on
board of the Infrared Space Observatory (ISO, Kessler et al.
1996). The ISO spectra we use were presented by Lenorzer et al.
(2002).

This paper is organized as follows: in Sect.~\ref{ISO} we introduce the
sample of stars and the observations; Sect.~\ref{diagnostic} discusses the line
fluxes and the \hu/\bra\ versus \hu/\pfg\ diagram. Sect.~\ref{conclusion} summarizes
the results of this paper.

\section{The sample of stars and the ISO spectra}
\label{ISO}
\begin{figure}[t]
\resizebox{8.5cm}{!}{\includegraphics{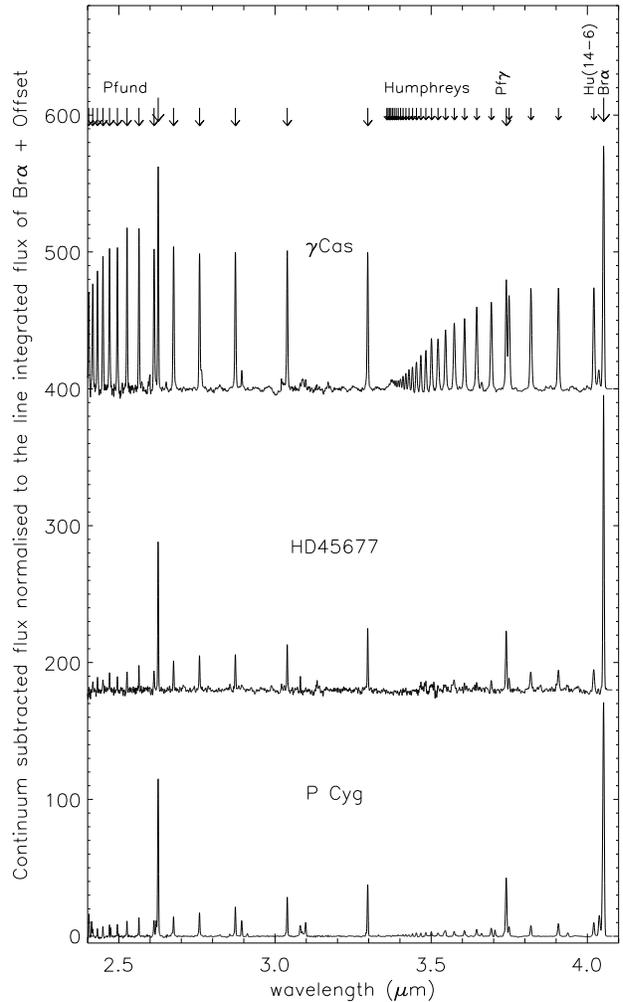}}  
\caption{Typical examples of the three classes of hot star
IR spectra considered. Top spectrum: the Be star $\gamma$~Cas; middle
spectrum: the B[e] star HD~45677; lower spectrum: the LBV P~Cygni 
(see text for discussion).}
\label{spectra}
\end{figure}

The ISO database contains infrared spectra of a wide range of hot
stars, observed in different programmes. In Lenorzer et al. (2002)
a homogeneous sample of ISO-SWS hot star spectra is presented, 
and line fluxes are derived. We use the results
of that study in the present analysis. All stars discussed here
were observed in at least the 2.4-4.1 $\mu$m spectral region
(band 1 of SWS; de Graauw et al. 1996). Fig.~\ref{spectra} displays 
representative spectra of the three classes of stars to which our
diagnostic can be applied (i.e. \hu\ can be measured): the Luminous
Blue Variables (LBVs), the B[e] stars and the Be stars. Before
discussing the ISO spectra, we briefly summarize the main
properties of these three classes of objects.

\emph{LBVs} are a rare class of hot massive post-main-sequence
stars characterized by a dense ionized stellar wind
($\dot{M}~\approx~10^{-5}~$M$_{\odot}/$yr) expanding at modest
speed (v$_{\rm exp}~=~100-300$~km/s) (Humphreys \& Davidson 1994).
The LBVs show variability on a range of timescales and amplitudes,
whose nature is not understood (e.g. van Genderen 2001). 
Recent high quality observations in the radio (Skinner et al. 1997) and 
in H$\alpha$ (Vakili et al. 1997 and Chesneau et al. 2000) show
that the wind of P~Cygni is variable and clumpy, however
its overall geometry does not strongly differ from spherical.
The near-IR spectrum of LBV's is dominated by free-bound and  
free-free emission from the wind, as
well as by strong recombination lines of H and He.
Many LBVs also show forbidden lines indicative of an extended 
low-density region. LBVs are known for the presence of extended, 
mostly dusty nebulae that are the result of large mass ejection in 
the (recent) past.

\emph{B[e]} stars are a poorly characterized class of hot stars
surrounded by large amounts of ionized and neutral/molecular
material. The luminosity class of B[e] stars can range from dwarf
to supergiant. At infrared wavelengths, a large excess due to the
presence of hot circumstellar dust is observed (e.g. Swings 1974); 
in addition, hot CO gas is detected in some stars
(MgGregor et al. 1989). B[e] stars have prominent \hi\ recombination
lines and strong forbidden lines. Spectropolarimetric observations
suggest that the circumstellar material is not distributed
spherically symmetric, but rather disk-like (see e.g. Bjorkman et al. 1998).

\emph{Be} stars are rapidly rotating B type dwarfs or giants that
show, or have shown, H$\alpha$ emission. The H$\alpha$ line is often
double-peaked and the line width scales with the projected rotational
velocity of the photosphere (see e.g. Dachs et al.  1986). The
infrared spectrum is dominated by a large excess due to ionized
circumstellar gas of high density, as well as by strong \hi\ (and, for
the hottest Be stars, \hei) recombination line emission.  Interferometric,
direct imaging as well as polarimetric observations show that the gas must be
located in a highly flattened circumstellar disk 
(e.g. Quirrenbach et al. 1997).  Be stars show no
circumstellar dust (with only a few exceptions) and also lack
molecular line emission and forbidden lines. These observations are
consistent with the presence of high-density gas in a flattened disk.

In Fig.~\ref{spectra} we show three ISO-SWS spectra, of the LBV
P~Cygni, of the B[e] star HD~45677, and of the Be star $\gamma$~Cas,
respectively (Lenorzer et al. 2002; see also Lamers et al. 1996a,b;
Malfait et al.  1998, and Hony et al. 2000 for detailed discussions of
the ISO data). The spectra have been continuum subtracted and normalized
to the line integrated flux of \bra\ to facilitate comparision. 
The three spectra show a remarkable range in line flux behaviour. 
While the LBV shows very large line flux
ratios between lines, the B[e] and Be stars do this to a much lesser
extent. These observations suggest that if the line spectrum is mostly
due to a stellar wind, the \hi\ line fluxes roughly scale 
with the Einstein coefficients; this is much less the case in the 
B[e] and Be star, where lines of very different intrinsic 
strength show similar line fluxes.
{This behaviour suggests that optical depths effects are 
important. The optical depth being dependent on the density squared, the
line ratios are mainly probing the density distribution 
of the circumstellar material. Moreover, in the optically thick case 
the line ratios are probing
the dependence in emitting surface with the wavelength which is also
a function of the density stratification. 
In addition, non-LTE effects may play a modest role,
as discussed by Hony et al. (2000).

{ It must therefore be possible to infer the density 
of the circumstellar material from the \hi\ IR 
recombination lines without resolving them.} 
In Sect.~\ref{diagnostic}, we will present a simple diagram which 
quantifies this density probe.

\section{Line flux ratio diagnostic}
\label{diagnostic}
\begin{figure}[t]
\resizebox{8.5cm}{!}{\includegraphics{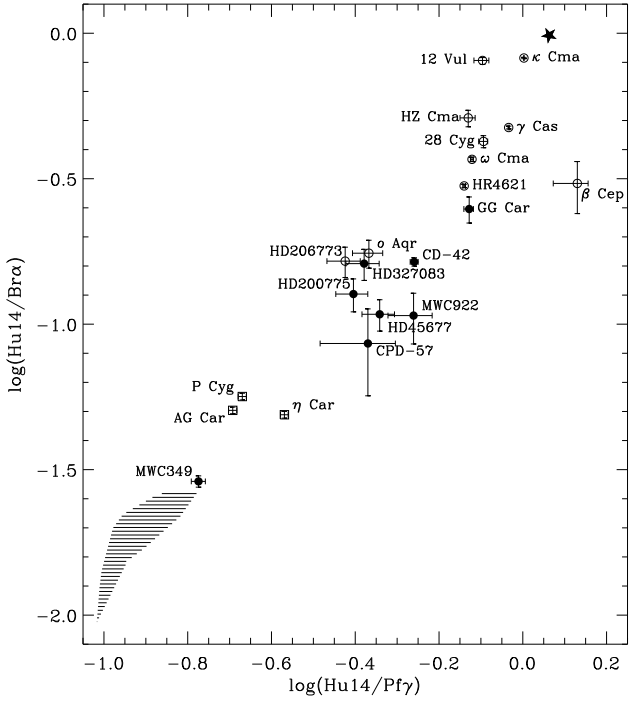}} 
\caption{\hu/\bra\ versus \hu/\pfg\ line ratio diagram for the
hot stars observed with ISO (Lenorzer et al. 2002). The different
classes of objects, LBVs (squares), B[e] stars (filled circles) and 
Be stars (open circles), are well separated. The thick asterisks 
indicates the position of optically thick black body emission; 
the striped region shows
the range of ratios for Menzel case B recombination, including
collisional de-excitation, for temperatures higher than $10^{4}$K.}
\label{diagram}
\end{figure}

Fig.~\ref{diagram} presents a diagram in which the line flux ratio
of \hu/\bra\ versus \hu/\pfg\ is plotted for different types of emission
line objects. The aim of this diagram is to provide a simple means to
investigate the nature of circumstellar gas in highly obscured sources.

The diagram shows a clear trend in that both line flux ratios typically
increase from LBV to B[e] to Be stars. This trend can be understood
in terms of the span in mass absorption coefficient between
\hu, \pfg\ and \bra. In an optically thin medium one expects the
plotted line flux ratios to follow Menzel Case B recombination theory.
However, in an optically thick medium the plotted ratio becomes
independent of mass absorption coefficient as the flux in any line
is dominated by the size of the emitting surface. 
For optically thick lines the line flux ratio of two lines can be 
written as:
\begin{equation}
\frac{I_1}{I_2} = \frac{B(\nu_1,T) S_{\rm eff,1}}{B(\nu_2,T) S_{\rm eff,2}}
\label{eq:thick}
\end{equation}
where B($\nu$,T) is the Planck function and S$_{\rm eff}$ is the
effective radiating surface in the line. If S$_{\rm eff}$ is similar
for both lines { and the Rayleigh-Jeans approximation is applied}, 
Eq.~\ref{eq:thick} reduces to: ${I_1}/{I_2} = {\nu_1^2}/{\nu_2^2}$.

This limit is indicated by a filled star in Fig.~\ref{diagram} 
and falls close to the locus of the Be stars. Note that we have 
assumed that the lines are in LTE and that the emitting medium is isothermal. 
Several Be stars deviate somewhat from the optically thick limit 
suggesting that there is a contribution from optically thin gas. 
This is to be expected since Be star disks can extend to 10-100 R$_*$ and, 
depending on the density gradient in the disk, optically thin 
gas must contribute. 
{ Inclination may also play a role in that disks seen edge-on are expected
to have a larger contribution from optically thin material than do
disks seen face-on.}
Finally, it is known that disks of Be stars are of transient nature, their
densities change with time, such that their position in the diagram 
is expected to vary more or less along the diagonal.

The locus of an optically thin isothermal gas emitting line radiation
according to Menzel case~B and for temperature greater than $10^{4}$ K 
is indicated by the dashed area in Fig.~\ref{diagram} (from
Storey \& Hummer 1995).
This region is close to that of the LBVs, suggesting that the
bulk of the line emission from these stars is due to optically thin
gas. This may be expected, since the winds are rapidly expanding and
the density decreases as r$^{-2}$ or steeper (in the innermost
regions).  The high mass loss rate in a roughly spherical outflow
causes a large emission measure and thus large line fluxes for the
strongest lines, and a detectable signal even for the intrinsically
much weaker lines.

The B[e] stars occupy a region in Fig.~\ref{diagram} which spans
in between most of the Be and the LBV stars. If our interpretation
of the line flux ratios in terms of densities is correct, this would
imply that the B[e] stars have contributions from both optically
thick and thin regions to the line flux. This is qualitatively 
consistent with the notion that B[e] stars have circumstellar disks
(with high densities) but that the presence of forbidden line emission
shows that an extended region of lower-density ionized gas must
also be present. Perhaps the scale-height of the disks surrounding B[e] 
stars is larger, or their disk radii are larger compared to the Be stars.

It is interesting to discuss the location of some well-studied,
enigmatic objects in Fig.~\ref{diagram}. The LBV $\eta$~Car is
located near the Menzel case~B limit, suggesting that the wind dominates the
line emission from this star, and not the extended nebula. Detailed
non-LTE model calculations indeed show that the UV to near-IR
spectrum of $\eta$~Car can be fitted well using a very dense wind
(Hillier et al. 2001). The peculiar star
MWC349 is also located near the case B limit, again suggesting that
the lines are formed in an optically thin medium. Polarimetry of MWC349 however
shows that a dusty disk is present (Yudin 1996), but this disk is not
contributing significantly to the emission from the ionized gas. 
This is consistent with the observed radio continuum spectrum of MWC349, 
which has the canonical $\nu^{0.6}$ slope expected for a spherically symmetric
constant velocity wind (Rodriguez et al. 1986). 

The example of MWC349 illustrates the care that needs to be taken 
when interpreting observations using our diagnostic diagram: { stars close 
to the Menzel case~B location may have a disk, however 
{\it the \hi\ lines are mainly not formed in the disk}.
The same care has to be taken when interpreting stars in the upper right 
corner of Fig.~\ref{diagram}, where the line flux ratios are close
to unity. For these, the conclusion must be that the
lines are opaque and that the geometry of the main \hi\ line emitting
region is likely a disk. However, from the line flux ratio's alone one
can not exclude geometries other than a disk, such as an optically thick
(expanding) shell. Note that one may likely distinguish between these 
different geometries using kinematic information from resolved line profiles.} 

In first order, both \pfg\ and \bra\ probe the
emission measure of the gas. This explains why the stars are more-or-less
concentrated around the diagonal. So, in principle one can do the
analysis using only one of the two line ratios. However, both lines
have a specific advantage. The pro of \bra\ is that it is the strongest
line, while the advantage of \pfg\ is that it suffers less from
possible contamination by nebular emission.
Plotting the line fluxes relative to \hu, the strongest Humphreys
series line in the L'-band, provides a better contrast
between contributions from optically thin and thick material than
does \bra\ relative to \pfg. We note that if reddening is
significant it will affect \hu/\pfg\ by shifting it to the right
in the diagram by about $0.07 \times \log ($\hu/\pfg$)$ per magnitude 
of extinction. \hu/\bra\ is not significantly affected as the 
wavelength separation between these two lines is very modest.

\section{Conclusion}
\label{conclusion}
A considerable fraction of the massive stars of our galaxy is obscured 
by dust. Dust extinction is preventing us from using powerful 
criteria, developed at UV and optical wavelength ranges, to infer 
physical properties of hot stars. At wavelengths longwards of about 5 microns,
emission from warm dust is typically dominating the spectra. 
It is therefore crucial 
to develop diagnostic tools in the near-infrared window in order to
understand the nature and evolution of hot stars surrounded by dust, as
it is the case, for instance, during the early stages of their lives in
giant molecular clouds. 
The diagram presented in Fig.~\ref{diagram} provides such a tool. 
It gives a simple means to use line fluxes observed in the 
infrared L' window to constrain aspects of 
the density and spatial distribution of circumstellar gas around hot stars
which may greatly help identifying the nature of such obscured sources.

\acknowledgements{We would like to thank S. Hony and C. Neiner 
for useful
discussions, and B. Vandenbussche and P. Morris for invaluable help in 
obtaining and analysing the ISO observations.
We acknowledge support from an NWO
'Pionier' grant. 
This work was partly supported by NWO Spinoza grant 08-0 to E.P.J. van den Heuvel.
}

\end{document}